\documentclass[useAMS,usenatbib,usegraphicx]{mn2e}

\usepackage{graphicx,times}
\usepackage[authoryear]{natbib} 
\bibpunct{(}{)}{,}{a}{}{,} 

\def\arcsec{\hbox{$^{\prime\prime}$}}
\def\ion#1#2{#1\,{\sc #2}}

\def\ion#1#2{#1\,{\sc #2}}
\def\aj{AJ}
\def\apj{ApJ}
\def\apjl{ApJ}
\def\apjs{ApJS}
\def\aap{A\&A}
\def\aaps{A\&AS}
\def\iaucirc{IAU~Circ.}
\def\mnras{MNRAS}
\def\pasp{PASP}
\title[$K$-band mini-survey of Galactic \protect{B[e]}
  stars] {A $K$-band spectral mini-survey of Galactic
  \protect{B[e]} stars\,\thanks{Based on data acquired using the Large
    Binocular Telescope (LBT) and Gemini Observatory. The LBT is an
    international collaboration among institutions in Germany, Italy,
    and the United States. LBT Corporation partners are LBT
    Beteiligungsgesellschaft, Germany, representing the Max Planck
    Society, the Astrophysical Institute Potsdam, and Heidelberg
    University; Istituto Nazionale di Astrofisica, Italy; The
    University of Arizona on behalf of the Arizona university system;
    The Ohio State University, and The Research Corporation, on behalf
    of the University of Notre Dame, University of Minnesota, and
    University of Virginia.  Gemini Observatory is operated by the
    Association of Universities for Research in Astronomy, Inc., under
    a cooperative agreement with the NSF on behalf of the Gemini
    partnership: the National Science Foundation (United States), the
    National Research Council (Canada), CONICYT (Chile), the
    Australian Research Council (Australia), Minist\'erio da
    Ci\v{e}ncia, Tecnologia e Inova\c{c}\~{a}o (Brazil) and Ministerio
    de Ciencia, Tecnolog\'{i}a e Innovaci\'on Productiva (Argentina).”
}}

\author[A. Liermann et al.]{A.~Liermann$^{1,2}$,
          O.~Schnurr$^{2}$,
          M.~Kraus$^{3}$
          A.~Kreplin$^{1}$
          M.~L.~Arias$^{4,5}$
          L.~S.~Cidale$^{4,5}$\\
$^{1}$ Max-Planck-Institut f\"ur Radioastronomie, Auf dem H\"ugel 69,
53121 Bonn, Germany\\
$^{2}$ Astrophysikalisches Institut Potsdam, An der Sternwarte 16, 14482
Potsdam, Germany\\
$^{3}$ Astronomick\'y \'ustav, Akademie v\v{e}d \v{C}esk\'e republiky,
Fri\v{c}ova 298, 251\,65 Ond\v{r}ejov, Czech Republic\\ 
$^{4}$ Departamento de Espectroscop\'ia Estelar, Facultad de Ciencias
Astron\'omicas y Geof\'isicas, Universidad Nacional de La Plata,\\ \quad Paseo
del Bosque s/n, B1900FWA, La Plata, Argentina\\
$^{5}$ Instituto de Astrof\'isica de La Plata, CCT La Plata,
CONICET-UNLP, Paseo del Bosque s/n, B1900FWA, La Plata, Argentina
}

\begin{document}

\date{Accepted 2014 June 12. Received 2014 June 11; in
  original form 2014 March 25}

\pagerange{\pageref{firstpage}--\pageref{lastpage}} \pubyear{2014}

\maketitle

\label{firstpage}

\begin{abstract}
  We present a mini-survey of Galactic B[e] stars mainly undertaken
  with the Large Binocular Telescope (LBT).  B[e] stars show
  morphological features with hydrogen emission lines and an infrared
  excess, attributed to warm circumstellar dust. In general, these
  features are assumed to arise from dense, non-spherical,
  disk-forming circumstellar material in which molecules and dust can
  condensate. Due to the lack of reliable luminosities, the class of
  Galactic B[e] stars contains stars at very different stellar
  evolutionary phases like Herbig AeBe, supergiants or planetary
  nebulae. \\ We took near-infrared long-slit $K$-band spectra for a
  sample of Galactic B[e] stars with the LBT-Luci\,I. Prominent
  spectral features, such as the Brackett\,$\gamma$ line and CO band
  heads are identified in the spectra. The analysis shows that the
  stars can be characterized as evolved objects. Among others we find
  one LBV candidate (MWC\,314), one supergiant B[e] candidate with
  $^{13}$CO (MWC\,137) and in two cases (MWC\,623 and AS\,381)
  indications for the existence of a late-type binary companion,
  complementary to previous studies.\\
  For MWC\,84, IR spectra were taken at different epochs with
  LBT-Luci\,I and the GNIRS spectrograph at the Gemini North
  telescope. The new data show the disappearance of the circumstellar
  CO emission around this star, previously detectable over decades. Also
  no signs of a recent prominent eruption leading to the formation of
  new CO disk emission are found during 2010 and 2013.

\end{abstract}


\section{Introduction}

B[e] stars are enigmatic objects. Their optical spectra show strong
Balmer emission lines as well as permitted and forbidden emission
lines of lowly-ionized metals. In addition, B[e] stars display a near-
and mid-infrared excess that is attributed to hot and warm
circumstellar dust. However, because the definition of the B[e] class
is purely morphological, it contains objects that are physically very
different in terms of their initial mass and evolutionary phase: B[e]
supergiants (B[e]SGs), Herbig AeBe (HAeBe) stars, compact planetary
nebulae, and symbiotic objects \citep[e.g.,][]{Lamers+1998}.
Especially B[e]SGs and HAeBe stars are difficult to distinguish; they
have dense, cool, and dusty equatorial disks, which are related to
either accretion and disk winds (in the case of HAeBe stars) or
equatorially enhanced stellar winds (B[e]SGs). These disks give rise
to molecular emission such as the first overtone bands of carbon
monoxide (CO) in the near-infrared which are observed in both HAeBe
stars and B[e]SGs \citep{Bik+2006, McGregor+1988, Morris+1996}. In
addition, the position of HAeBe stars in the empirical
Hertzsprung-Russell diagram overlaps with that of the low-luminosity
B[e]SGs.

So far, comprehensive studies of B[e] stars, focused on the spectral
classification and characterization, were mostly based on optical
spectra. Over the last two decades new instruments have given access
to the infrared (IR) spectral range and allowed high-quality spectra
with the necessary spectral resolution to be obtained. But only now
projects are checking systematically how appropriate spectral
classification based on IR spectra alone is for general application
purposes.  For example, \citet{Oksala+2013} find that based on
$K$-band spectra three distinct groups of stars can be identified: (1)
``regular'' B[e]SGs with the expected spectrum of emission lines
including the Pfund series and CO in emission, (2) S\,Dor-like luminous
blue variables (LBVs) with a variety of strong emission lines but
lacking the expected circumstellar CO emission, and (3) a group of cool
stars, consisting of LBVs in outburst and Yellow Hypergiants (YHGs),
showing the Pfund series in absorption.

The CO bands can be used to obtain an age estimate for a star. As
proposed by \citet{Kraus2009}, stellar evolution models with rotation
predict the enhancement of the carbon isotope $^{13}$C on the stellar
surface during the core-hydrogen burning of massive stars through
mixing. Via mass loss, it is transported into the circumstellar
environments and locked into $^{13}$CO molecules. Hence, in evolved stars
(i.e., supergiants), the enriched isotope should become detectable as
$^{13}$CO bands. This was confirmed by detection of $^{13}$CO emission
in known, extra-galactic B[e]SGs \citep{Liermann+2010}; vice versa the
$^{13}$CO absorption can be used to distinguish late-type supergiants
from dwarf stars \citep{Wallace-Hinkle1997}.

We have embarked on a mini-survey of a sample of Galactic B[e] stars
to investigate their evolutionary status and characterize their age
and circumstellar material from $K$-band spectra.  We present our
observations in Sect.\,\ref{sec:obs} with the results of the spectral
analysis following in Sect.\,\ref{sec:results}. The final part of the
paper is dedicated to the discussion and conclusion of our findings
(Sects.\,\ref{sec:discussion} and \ref{sec:conclusion}).

\section{Observation and data reduction}
\label{sec:obs}

   \begin{table}
      \caption{Log of the LBT observations.}
      \label{tab:obs}
         \begin{tabular}{@{}lcccrccc}
            \hline 
Object      & RA & DEC & Date (UT) & $t_{\rm int}$ [s] \\
            \hline 
Science: &\\ 
MWC\,137    & 06 18 45.52& +15 16 52.25& 2010-11-10 &  900  \\
MWC\,84     & 04 19 42.14& +55 59 57.70& 2010-11-10 &  120  \\
MWC\,314    & 19 21 33.97& +14 52 56.89& 2010-11-13 &  240  \\
AS\,381  & 20 06 39.95& +33 14 28.10& 2010-11-13 &  600  \\
MWC\,300    & 18 29 25.69& -06 04 37.29& 2011-05-12 &  200  \\
MWC\,623    & 19 56 31.54& +31 06 20.12& 2011-05-12 &   80  \\
            \hline 
Standards:  &\\ 
HD\,44585   & 06 23 09.15& +15 50 32.33& 2010-11-10 & 1440  \\
HD\,29371   & 04 40 49.54& +57 52 46.63& 2010-11-10 &  600  \\
HD\,185195  & 19 37 17.84& +15 15 02.41& 2010-11-13 &  900  \\
HD\,196006  & 20 33 30.39& +32 54 27.78& 2010-11-13 &  900  \\
HD\,175644  & 18 56 47.20& -14 01 40.76& 2011-05-12 &  400  \\
HD\,191720  & 20 10 01.65& +36 58 42.47& 2011-05-12 &  400  \\
            \hline 
         \end{tabular}
   \end{table}

Our sample stars (cf. Table\,\ref{tab:obs}) were observed with the
Large Binocular Telescope (LBT) in Arizona, USA, during 3 runs in
November 2010, April and May 2011.  We used the LBT-Luci\,I
spectrograph \citep{Seifert+2003} to obtain seeing-limited,
high-quality, long-slit spectra with the N1.8 camera, a slit width of
0.5\,$\arcsec$, and the $K$-band filter (1.93--2.48\,$\mu$m) with the
210\_zJHK grating tilted to center on $\lambda = 2.24\,\mu$m to have
best wavelength coverage for the CO bands redwards of 2.29\,$\mu$m.
The resulting spectra have a spectral resolution of about $R = $6,100
and a signal to noise ratio of S/N$\approx$700.

Telluric standards were observed immediately before or after the
science targets at similar airmass. B main-sequence stars were
used as standard stars, to avoid the contamination of the CO bands
with intrinsic features from a B supergiant or solar-type standard
star. Table\,\ref{tab:obs} lists the details of the
observations.

Additional data for MWC\,84 were obtained during April, 2011 and
October, 2013 with GEMINI/GNIRS (programs GN-2011B-Q-24 and
GN-2013B-Q-11) in high-resolution mode ($R$=18,000), using the 110.5
l/mm grating, the long camera (0.05\,$\arcsec$/pix), and the
0.10\,$\arcsec$ slit. We obtained spectra in the $K$-band
(2.28--2.34\,$\mu$m) centered on $\lambda = 2.31\,\mu$m to cover the
first two CO bandheads.

Data sets from both telescopes were reduced using standard {\sc
  iraf}\footnote{IRAF is distributed by the National Optical Astronomy
  Observatories, which are operated by the Association of Universities
  for Research in Astronomy, Inc., under cooperative agreement with
  the National Science Foundation.} routines for long-slit
spectroscopy including removal of cosmic rays and dead pixels, dark frames,
division by a normalized flat field, distortion correction, and
wavelength calibration. Sky subtraction was done by subtracting offset
frames per target (AB pattern) correspondingly before the spectra
extraction.

The frames of the telluric standard stars were treated in the same
way. We produced calibration curves as the ratio of
the extracted standard star spectra and Kurucz models
\citep[corresponding to the spectral types of the standard
  stars,][]{Kurucz1993} scaled to the stars' 2MASS $K$-band
magnitudes. As the Kurucz models do not reproduce the observed
Br\,$\gamma$ line in the standard star spectra that part of the
calibration curve was smoothed by linear interpolation between 2.15
and 2.17\,$\mu$m.
For flux calibration, the extracted science target spectra were divided
by the corresponding calibration curves.

\section{Analysis and Results}
\label{sec:results}

\begin{figure*}
 \includegraphics[width=0.95\textwidth]{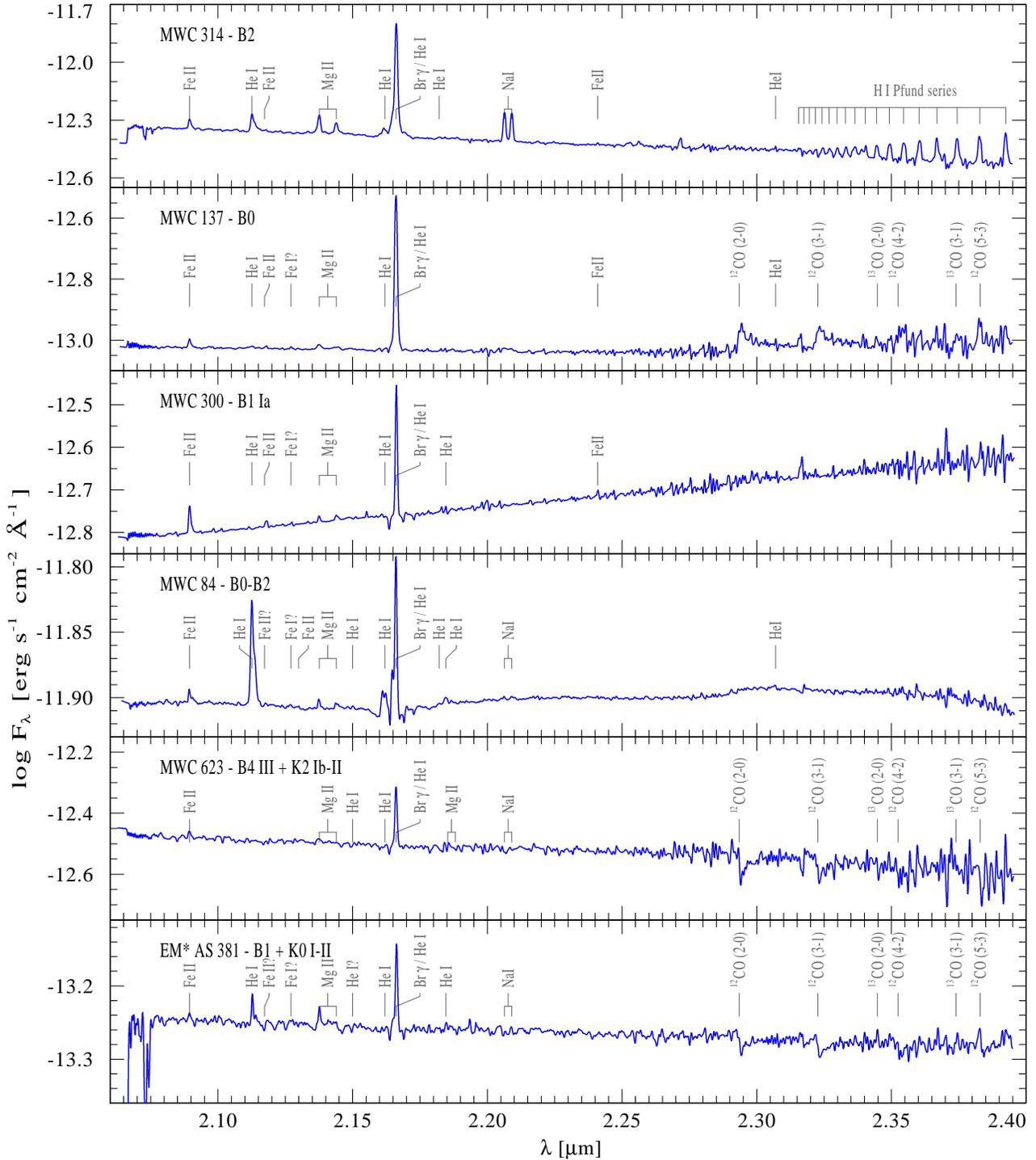} 
\caption{Flux-calibrated $K$-band spectra of our sample stars observed
  with LBT-Luci\,I. Line identifications of prominent emission lines
  and the band heads of $^{12}$CO and $^{13}$CO are indicated. Please
  note that the scaling of the flux axis is different for each
  panel. Spectral types listed for the individual stars are taken from
  the literature as listed in Table\,\ref{tab:starparameters}.}
  \label{fig:spectra}
\end{figure*}

   \begin{table*}
      \caption{Equivalent width measurements (in \AA\,) for prominent
        emission lines (negative) and absorption lines
        (positive).}
      \label{tab:linelist}
         \begin{tabular}{lccccccccccc}
            \hline 
star & Fe\,II& He\,I& \multicolumn{2}{c}{Mg\,II}& Br\,$\gamma$& \multicolumn{2}{c}{Na\,I}& CO (2-0)& CO (3-1)\\
     & 2.089\,$\mu$m& 2.112\,$\mu$m& \multicolumn{2}{c}{2.138/144\,$\mu$m}& 2.166\,$\mu$m
&\multicolumn{2}{c}{2.206/209\,$\mu$m}& 2.293\,$\mu$m& 2.322\,$\mu$m\\ 
\hline 
MWC\,314     & -2.0 & -3.6 & -3.3&-2.7 & -32.1 & -5.1&-5.0 & -     & -\\
MWC\,137     & -0.9 & -    & -0.8&-0.4 & -25.6 & -   & -   & -21.1 & -19.1\\
MWC\,300     & -1.9 & -    & -0.3&-0.5 & -8.7  & -   & -   & -     & -\\
MWC\,84      & -0.4 & -3.4 & -0.3&-0.4 & -4.3  & -0.1&-0.2 & -     & -\\
MWC\,623     & -0.9 & -    & -0.8& -   & -5.8  & -   & -   & 14.0  & 11.6 \\
AS\,381      & -0.5 & -0.8 & -1.2&-1.3 & -3.6  & -0.6&-0.4 & 5.5   & 5.3 \\
\hline    
         \end{tabular}
   \end{table*}

The flux-calibrated spectra of the sample stars are presented in
Fig.\,\ref{fig:spectra}. Most spectra show more or less pronounced
telluric residuals at about 2.316 and 2.370\,$\mu$m.  We attribute
this to the not always perfect match of observing conditions between
the science target and standard star.

For all sample stars we clearly detect the Brackett\,$\gamma$ line
(Br\,$\gamma$, $\lambda$\,2.166\,$\mu$m) in emission.  Additionally,
iron (\ion{Fe}{ii} $\lambda$\,2.089\,$\mu$m) and a magnesium doublet
(\ion{Mg}{ii} $\lambda$\,2.138/144\,$\mu$m) are detected. Three stars,
MWC\,314, MWC\,84, and AS\,381, also show features of the sodium
doublet (\ion{Na}{i} $\lambda$\,2.206/209\,$\mu$m) and helium
(\ion{He}{i} $\lambda$\,2.112\,$\mu$m) in emission.  In
Table\,\ref{tab:linelist} we list the measured equivalent widths of
these lines for all stars.

Remarkably, MWC\,314 is the only star whose spectrum shows very
prominently lines of the Pfund series (see Fig.\,\ref{fig:spectra},
top-most panel). Overall its spectral appearance is close to
S-Dor like Luminous Blue Variables (LBVs) and B[e]SGs
\citep{Oksala+2013}; a classification of MWC\,314 as LBV will be
discussed below (see Sect.\,\ref{sec:mwc314}). Also, no molecular CO
bands, neither in emission nor in absorption, are present in the
spectrum.

\begin{figure}
 \includegraphics[width=\columnwidth]{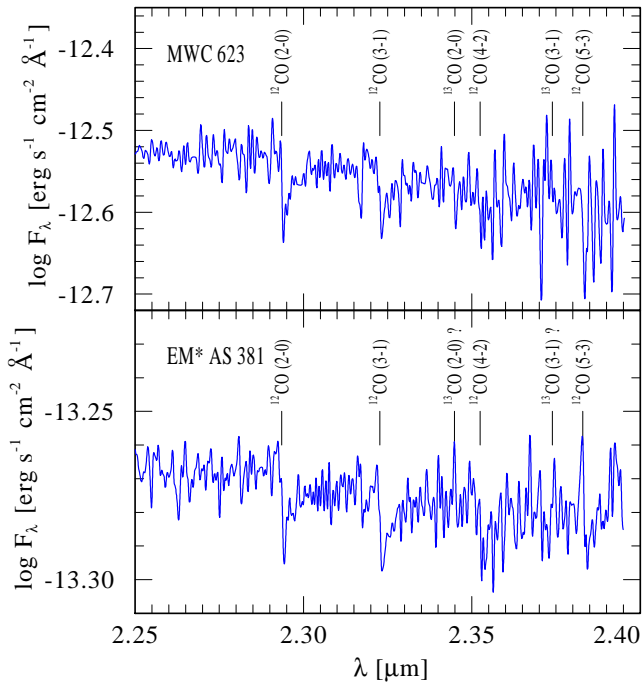}
\caption{Detail of the spectra of MWC\,623 and AS\,381 showing the CO
  absorption attributed to their cool companions.}
  \label{fig:mwc623-as381-co}
\end{figure}

However, CO bands are detected in the spectra of three stars: MWC\,137
(in emission), MWC\,623 and AS\,381 (in absorption). Details of the
spectra are shown in Figs.\,\ref{fig:mwc137-co} and
\,\ref{fig:mwc623-as381-co}.  MWC\,137 displays $^{13}$CO in emission
which shows that the star is evolved, i.e. possibly a B[e]SG.

Both MWC\,623 and AS\,381 have been suspected to be binary stars; for
MWC\,623 \citet{Zickgraf-Stahl1989} find a spectroscopic binary with
two sets of spectral lines (SB2) and for AS\,381 \citet{Miro+2002b}
find neutral metal lines inferring a cool companion.  Indeed, the
spectra of both stars display CO absorptions which we attribute to a
cool late-type companion, respectively. We measure the equivalent
widths of the first band head (CO 2-0), see
Table\,\ref{tab:linelist}. Following \citet[see Eqs.\,(5) and
(6)]{Gonzalez-Fernandez+2008}, the equivalent width can be used to
derive both the effective temperature and a spectral-type of the
companion (the indicator $G =$ 0 \ldots 5 \ldots 13 corresponds to
spectral types K0 \ldots K5 \ldots M7).

For MWC\,623 we find a K4 companion ($G=$ 3.2 to 5.1 corresponding to
spectral types between K3 to K5) with an effective temperature of
$T_\mathrm{eff} = 4030 \pm 100$\,K.  In the case of AS\,381, we
determine a spectral type of the companion of K0 ($G=$ 0.09 to 0.20
corresponding to a spectral type K0.)  with an effective temperature
of $T_\mathrm{eff} = 4550 \pm 100$\,K.  Errors on the effective
temperature are dominated by those given in Eq.\,5
of \citet{Gonzalez-Fernandez+2008}.  In both cases one can argue for
the detection of weak $^{13}$CO absorptions in the spectra, indicating
a slightly evolved giant or supergiant companion (luminosity class of
II to I).

Both MWC\,84 and MWC\,300 do not show any Pfund lines or CO emission
in their spectra. However, MWC\,84 shows the strongest \ion{He}{i} and
MWC\,300 the strongest \ion{Fe}{ii} and Br\,$\gamma$ lines in
emission, of all the sample stars.

\section{Discussion}
\label{sec:discussion}
\subsection{MWC\,314 -- a quiescent B[e]SG/LBV?}
\label{sec:mwc314}
For high-mass stars, stellar evolution models predict a short
transitional phase as LBV from the core hydrogen-burning OB star
progenitors to the core helium-burning Wolf-Rayet stars
\citep[e.g.][]{Maeder+2008}.  During the LBV phase the stars undergo
extreme mass loss events (``outbursts'') after or followed by a rather
stable (``quiescent'' or ``dormant'') state and are often associated
with circumstellar nebulae. In addition, significant variability in
brightness and spectral appearance can be detected more readily in
the outburst phase but also are present in the quiescent state.

For MWC\,314 spectral, photometric, and polarimetric variability was
found covering different periods, e.g. \citet{Miro1996,
  Wisniewski+2006, Groh+2007}. A detailed spectral analysis by
\citet{Miro+1998}, finding photospheric lines for the first time,
classified the star as B0 supergiant with $T_\mathrm{eff} =
25,000\,$K. Additionally, the authors present indications for a
non-spherical wind and derive a distance of $d= 3.0\pm
0.2\,$kpc from radial velocity ($RV$) measurements. This makes
MWC\,314 one of the most luminous stars in the Milky Way with
$\log{(L/L_\odot)}= 6.1 \pm 0.3$ \citep{Miro+1998}.

Later studies find different (effective) temperatures, based
on different indicators and methods, e.g,\ 26,700\,K to 32,000\,K
\citep{Cidale+2001}, 16,200\,K \citep{Carmona+2010}, 18,000\,K
\citep{Lobel+2013}. At the same time, different studies report the
rather stable photometric appearance of MWC\,314 in $UBVRIJHK$ filter
observations with about $0.3\,$mag variability, e.g.\ \citet[covering
the time span from 1954 to 1995]{Bergner+1995, Miro1996}. This results
in a conflicting situation to determine the spectral type and
evolutionary state of MWC\,314.

\begin{figure}
 \includegraphics[width=\columnwidth]{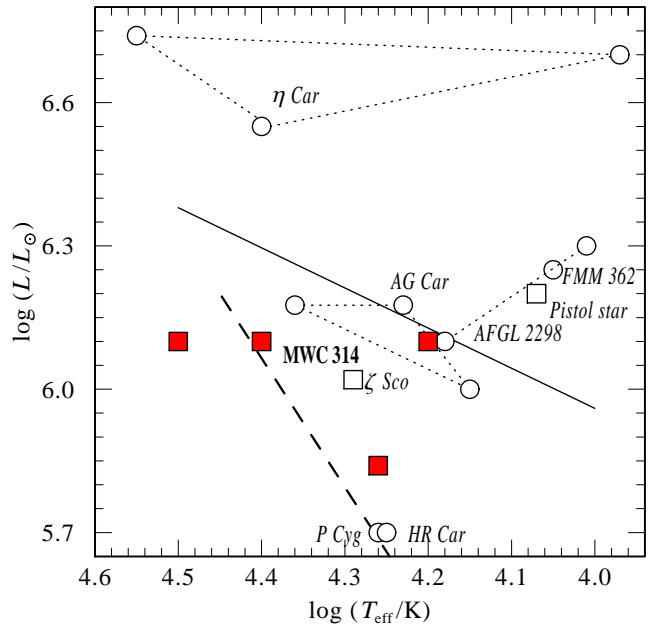}
\caption{HRD for known Galactic LBV stars (circles) and LBV candidates
  (squares).  The empirical Humphreys-Davidson limit \citep[solid
    line]{Humphreys-Davidson1994} and the hot LBV minimum light strip
  \citep[dashed line]{Clark+2005} are indicated for illustration. For
  MWC\,314, we show a range of fundamental parameters listed in
  Table\,\ref{tab:lbv}; for temperatures with no simultaneous
  luminosity determinations we adopt $\log{(L/L_\odot) =6.1}$ from
  \citet{Miro+1998}.}
  \label{fig:lbvs}
\end{figure}

The star was first proposed as LBV candidate by \citet{Miro1996}.  In
case the above listed temperatures reflect a real change over the last
20\,years, MWC\,314 would be rather similar to other LBVs (see
Table\,\ref{tab:lbv}), thus supporting the classification of MWC\,314
as LBV candidate. In Fig.\,\ref{fig:lbvs}, we show MWC\,314 in the
Hertzsprung-Russell diagram (HRD) among known Galactic LBVs and LBV
candidates. It is located between the empirical Humphreys-Davidson
limit \citep[][solid line in Fig.\,\ref{fig:lbvs}]{Humphreys-Davidson1994}
and the hot temperature LBV minimum light strip \citep[][dashed line
in Fig.\,\ref{fig:lbvs}]{Clark+2005}.

In our spectrum of MWC\,314 we find pronounced lines of
hydrogen, Br\,$\gamma$ and Pfund series, and prominent emission
lines of \ion{Fe}{ii}, \ion{Mg}{ii}, and \ion{Na}{i}.  Comparing the
spectra of MWC\,84, MWC\,137, and MWC\,314, we find most pronounced
\ion{Na}{i} and \ion{Mg}{ii} emission for the latter with remarkable
line strengths. Typically, comparable line strengths are only found in
late-type (F and G) supergiants \citep{Hanson+1996,
Wallace-Hinkle1997}. In B-type emission-line stars these lines are
usually (much) weaker, but with a slight trend for an increase towards
later types \citep[see][]{Oksala+2013}. If we consider that the
temperature obtained by \citet{Lobel+2013} within the past years is
the currently most reliable one, then we can assign MWC\,314 a
spectral type of B2 according to the temperature-spectral type
relation for Galactic supergiant stars of \citet{Searle+2008}.

Previous studies by \cite{Muratorio+2008} found indications for a
quasi-Keplerian\footnote{A disk in Keplerian rotation displaying
simultaneously a small outflow component, see
e.g., \citet{Krticka+2011, Kurfuerst+2013}.} circumstellar disk from
double-peaked emission lines. We do not detect any significant IR
excess that one would expect from warm/hot dust in the circumstellar
disk; this result is highly indicative for the absence of such
dust. We also do not find any CO emissions or absorptions. This is
comparable to other LBV candidates where the lack of CO emission has
been attributed to the circumstellar material having too low a
density \citep{Morris+1996, Oksala+2013}.

The presence of a dense and compact gas disk in MWC\,314 can
be concluded from the detection of double peaked [\ion{Ca}{ii}]
emission together with the non-detection of [\ion{O}{i}] by Aret et
al. (in preparation). 
\citet{Aret+2012} show, that [\ion{Ca}{ii}] traces
regions of higher densities, i.e. closer to the star. Thus the lack of
[\ion{O}{i}] in MWC\,314 must be attributed to a dense but compact
disk. On the other hand, the presence of the strong emission from
neutral sodium, which has a very low ionization potential, indicates
the existence of a second very dense, but cool region of circumstellar
material. To allow for the emission of \ion{Na}{i}, that region should
be shielded from direct stellar radiation. In a spherically-symmetric
stellar wind, too high densities would be required. But in such a high
density environment the emission of [\ion{O}{i}] can be
expected. Hence, it seems more logical to assume that the \ion{Na}{i}
emission arises from a cool circumstellar disk (or ring), which
provides ideal shielding conditions \citep[e.g.\
,][]{Scoville+1983,McGregor+1988a, McGregor+1988}.  However, at this
point we cannot exclude the option that the \ion{Na}{i} emission
arises from the environment of the companion.

\citet{Oksala+2013} present the $K$-band 
spectrum of LHA\,120-S\,127 and S\,Dor which look remarkably similar
to MWC\,314, finding disk tracers and no CO emission. The authors
state that CO molecules should form between the locations traced by
the Ca and O emitting material and dust particles, and conclude that
the disk of S\,127 might not be a continuous disk but rather
consisting of rings/shells. This cannot be explained with the model of
a continuous B[e] wind but rather requires an LBV eruption
scenario. Hence, for MWC\,314 we can assume a similar scenario, i.e.,
the presence of at least two different rings, a hot and compact one
close to the star from which the [CaII] lines arise, and very cool one
at much larger distance, where the NaI lines are excited. The
intermediate region must be of very low density due to the lack of
both [OI] and CO band emission.

%
   \begin{table}
\caption{Comparison of stellar parameters for
      LBV stars.}  \label{tab:lbv} \begin{tabular}{@{}lcccrccc} \hline
      star & $T_\mathrm{eff}$ [K]& $\log{L/L_\odot}$ &
      reference\\ \hline 
MWC\,314 (1986/91) & 25\,000 & 6.10 & (1) \\
MWC\,314 (1997/98) & 32\,000 & - & (2) \\ MWC\,314 (2007) & 16\,200 & - & (3) \\
MWC\,314 (2009-2011)      & 18\,000 & 5.84 & (13) \\
$\eta$\,Car               & 25\,000 & 6.57 & (4) \\
$\eta$\,Car               &  9\,400 & 6.70 & (11) \\
$\eta$\,Car               & 35\,300 & 6.74 & (12) \\
AG\,Car (min 1985-1990)   & 22\,800 & 6.17 & (5) \\
AG\,Car (min 2000-2001)   & 17\,000 & 6.17 & (5) \\
AG\,Car (max 2002-2003)   & 14\,000 & 6.00 & (6) \\
FMM\,362                  & 11\,300 & 6.25 & (7) \\
Pistol star               & 11\,800 & 6.20 & (7) \\
AFGL\,2298 (min 2006)     & 10\,300 & 6.30 & (14) \\
AFGL\,2298 (max 2001)     & 15\,000 & 6.10 & (8) \\
$\zeta$\,Sco              & 19\,500 & 6.02 & (9) \\
P\,Cyg (min 1980-2000)    & 18\,200 & 5.70 & (9) \\
HR\,Car                   & 17\,900 & 5.70 & (10) \\
            \hline 
         \end{tabular}

\smallskip
\raggedright{
  (1)~\citet{Miro+1998}, 
  (2)~\citet{Cidale+2001}, (3)~\citet{Carmona+2010},
  (4)~\citet{Humphreys-Davidson1994}, (5)~\citet{Groh+2011},
  (6)~\citet{Groh+2009b}, (7)~\citet{Najarro+2009}, (8)~\citet{Clark+2003},
  (9)~\citet{vanGenderen2001}, (10)~\citet{Groh+2009}, (11)~\citet{Groh+2012}, (12)~\citet{Hillier+2001}, (13)~\citet{Lobel+2013}, (14)~\citet{Clark+2009}
}
   \end{table}

A study by \citet{Marston-McCollum2008} found a bipolar nebula around
MWC\,314 that is similar in morphology to the one around
$\eta$\,Car. Their results are not conclusive whether or not that
nebula was ejected in a past LBV outburst or not.

Recently, it has been discussed for a large fraction of the known LBV
stars, that bipolar nebulae might be linked to a possible binary
nature of these stars. For MWC\,314 the binary status is still under
debate. Early reports of photometric variations \citep{Miro1996} were
interpreted as being more consistent with the pulsations of a slightly
evolved supergiant (or LBV candidate) rather than being attributed to
a binary companion. However, \citet{Muratorio+2008} find indications
for binarity based on $RV$ variations with an orbital
period of $P = 30.7\,$d. More recent studies continue the ambiguity of
the binary status, e.g. \citet{Rossi+2011} cannot detect any periodic
$RV$ variations. However, \citet{Lobel+2013} find
variations with a period of $P = 60.7\,$d (and an orbital eccentricity
of $e = 0.26$), inferring a massive, possibly evolved supergiant
companion to MWC\,314.  Our spectrum does not show any CO absorption
indicative for cool companions. Thus K and M-type companions can be
excluded for MWC\,314. However, a massive companion as suggested by
\cite{Lobel+2013} cannot be ruled out. Further high-resolution
spectroscopy and/or spatially-resolved imaging is necessary to confirm
or contradict the binary suspicion.

\subsection{MWC\,137 - a B[e]SG candidate with $^{13}$CO emission}

\begin{figure}
 \includegraphics[width=\columnwidth]{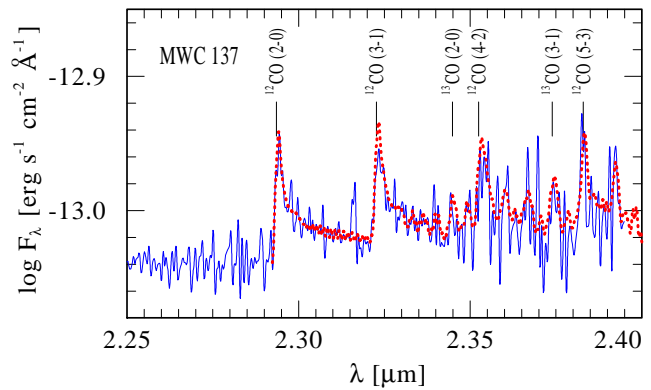}
\caption{Detail of the spectrum of MWC\,137 (solid line) showing the CO
  emission. The over-plotted model (dotted line) combines CO and Pfund
  emission added to the observed continuum.}
  \label{fig:mwc137-co}
\end{figure}

\begin{figure}
 \includegraphics[width=\columnwidth]{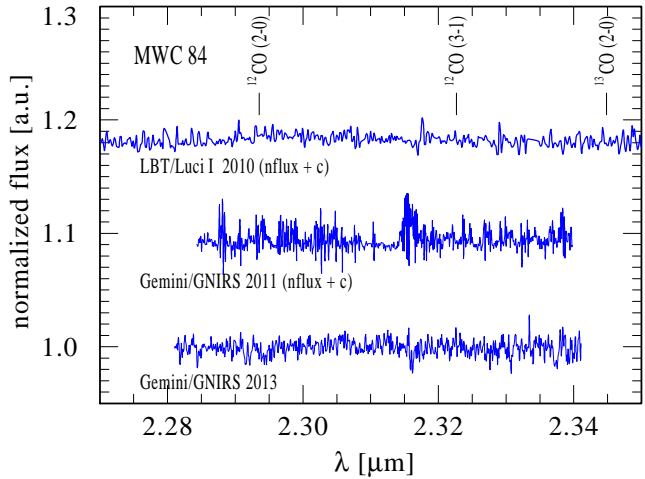}
\caption{Comparison of the spectra of MWC\,84 taken at the LBT and
  Gemini; spectra are shifted by a constant for better viewing. Over
  the covered time span of 3 years no CO emission is detected.}
  \label{fig:mwc084-co}
\end{figure}

This star is a typical example for the difficulty to distinguish
B[e]SG and HAeBe stars. Based on the preferred distance estimate, the
star has been considered to be a young unevolved object, e.g.\, with a low luminosity around $\log{(L/L_\odot)} = 4.4$,
assuming $d=1.3\,$kpc \citep{Hillenbrand+1992}. Also,
  \citet{Testi+1997} report the detection of an embedded cluster
  around MWC\,137 supporting the young nature of this object.
However, \citet{Esteban-Fernandez1998} derived $T_\mathrm{eff} =
30,000$\,K and $\log{(L/L_\odot)} = 5.37$ based on a re-assessment of
the distance to a lower limit of $d=6\,$kpc, and concluded that the
star should be classified as B[e]SG. In addition the authors argue
that the small photometric variations but stable spectral line
profiles found for MWC\,137, indicate the star being a B[e]SG rather
than a HAeBe star \citep[see also][]{Zickgraf1992}.  Moreover,
\citet{Esteban-Fernandez1998} find the associated ring nebula S\,226
around MWC\,137 to be isolated and not attached to any large-scale
star-forming region as might be expected for an HAeBe star. In
contrast these authors remark the spectroscopical resemblance of the
nebula to those observed around LBVs or Wolf-Rayet stars \citep[and
  references therein]{Esteban-Fernandez1998}.

\begin{figure}
 \includegraphics[width=\columnwidth]{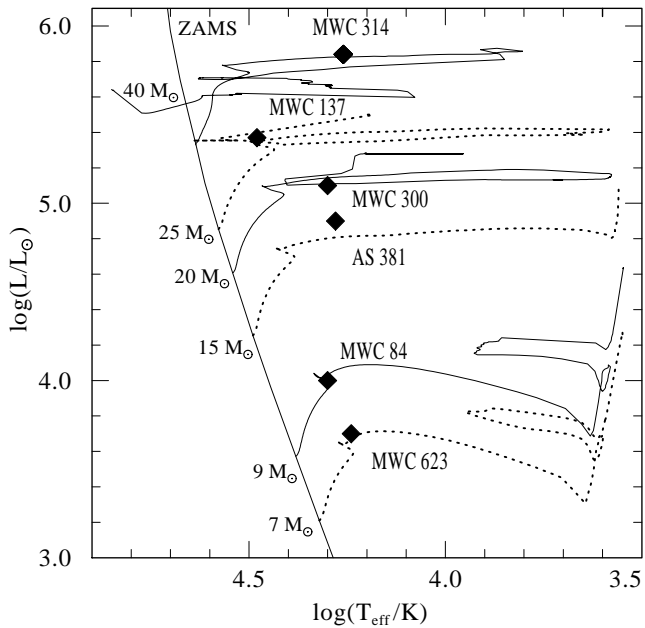}
\caption{HRD with our sample stars (stellar parameters listed in
  Table\,\ref{tab:starparameters}) and single star
  stellar evolution models for different initial masses by
  \citet{Ekstroem+2012}. The models account for the effects of stellar
  rotation.}
  \label{fig:hrd}
\end{figure}

\citet{Marston-McCollum2008} confirm the ring nebula and in addition
find a bipolar structure in their narrow-band images. However, the
nebula material seems chemically unprocessed which might indicate that
the circumstellar material either is swept-up interstellar matter or was
ejected during an early stellar-evolution phase.

MWC\,137 was listed by \citet{Miro2007} as FS\,CMa candidate star
which implies that it is a binary system.  However, neither
\citet{Baines+2006} nor \citet{Wheelwright+2010} found evidence for
the binary status of MWC\,137 from their spectro-astrometric
observations \citep[a method sensitive down to about 100\,mas binary
  separation, e.g.,][]{Bailey1998}.  \textsc{hipparcos} data
  suggest that MWC\,137 is an astrometric binary
  \citep{Makarov-Kaplan2005}, possibly with a white dwarf or subdwarf
  companion \citep{Lanning-Lepine2006}.

In our spectrum (see Fig.\,\ref{fig:spectra}), we do not find
indications for a cool companion in terms of CO absorption or a set of
emission or absorption lines accountable to a hot main-sequence
companion; instead, a closer look at our spectrum shows prominent CO
bandheads in emission, both of $^{12}$CO and $^{13}$CO.  In
Fig.\,\ref{fig:mwc137-co} we over-plot a model combining CO and Pfund
emission added to the observed continuum. The model was computed using
the codes of \citet{Kraus+2000, Kraus2009} and \citet{Oksala+2013};
parameters of the Pfund and CO emitting regions as obtained by
\citet{Oksala+2013}. The excellent agreement of the model and the
observations confirms the results by \citet{Oksala+2013}. In
particular the presence of clearly detectable emission from $^{13}$CO
implies that the circumstellar material is enriched in the $^{13}$C
isotope. This excludes a pre-main sequence nature according to the
stellar evolution models by \citet{Ekstroem+2012} and MWC\,137 has to
be classified as evolved, post-main sequence object.

\subsection{B[e] supergiant candidates}
In two further stars of our sample, MWC\,300 and MWC\,84, we do not
detect CO emission in their spectra. However, in the following we 
summarize the evidences for their classification as B[e]SG:

\paragraph*{MWC\,300} 
According to \citet{Appenzeller1977} this star is classified as
B1\,Ia. \citet{Miro+2004} detect photospheric lines and their spectral
analysis indicates the supergiant status of the star, with
$T_\mathrm{eff} = 20,000\,$K. From comparison of equivalent-width
measurements they derive a luminosity of $\log{(L/L_\odot)} =5.1 \pm
0.1$ assuming a distance of $d = 1.8 \pm 0.2$\,kpc. $RV$ variations
\citep{Miro+2004} and a clear signal from spectro-astrometry
\citep{Takami+2003} suggest that MWC\,300 is a binary. From
near-infrared interferometry, \citet{Wang+2012} find evidence for a
dusty circumstellar (maybe even circumbinary) disk and a binary
companion at a projected separation of about 7.9\,AU. No extended
circumstellar material has been detected in H\,$\alpha$ imaging by
\citet{Marston-McCollum2008}.

Our spectrum shows a clear infrared excess that we attribute to the
circumstellar/circumbinary dust, but we do not detect CO emission. As
for MWC\,314, this might indicate that the material's density is too
low to give rise to molecular emission.  In addition, we don't find
evidence for a cool companion by the lack of CO absorption expected
for late-type stars. As \citet{Miro+2004} suggests a companion of
about 6\,$M_\odot$, i.e. a B-type star, we scanned the
spectrum for the expected hydrogen absorption lines with no
result. This might indicate a supergiant companion for which the
spectral lines are less pronounced, such that basically only a
contribution to the continuum flux might be considered. Interestingly,
models of \citet{Wang+2012}, fitted to their interferometric data,
suggest the two binary components to be about similar in effective
temperature and to have a brightness ratio of about 2.2.

   \begin{table*}
      \caption{Stellar parameters for the sample stars.}
      \label{tab:starparameters}
         \begin{tabular}{llccccccll}
            \hline 
star &  SpT &$T_\mathrm{eff}$& $\log{(L/L_\odot)}$& $d$& $E(B-V)$&
$M_\mathrm{ini}$& age$^a$ & reference\\
     &      & [K]&  & [kpc]& [mag]& [$M_\odot$]& [Myr] & & \\
          \hline 
MWC\,314   & \textit{B2} & 18000& 5.84& 1.5$\pm$0.4& 1.45$\pm$0.15 & \textit{40} & \textit{6}& (1),(8)\\
MWC\,137   & B0 & 30000& 5.37 & $\ge$6 & 1.22 & 30, {\it 25} & {\it 8.0-8.4}& (2)\\
MWC\,300  & B1\,Ia & 20000 &5.1$\pm$0.1 & 1.8$\pm$0.2&
0.84$\pm$0.02&20, {\it 20}& {\it 9.5-10.3}& (7)\\
MWC\,84  & B0-2& 20000$\pm$2000& $<$ 4.0 & 2.2 & 0.85$\pm$0.05 &{\it 9} &
{\it 29.6-31.3} & (5), (6)\\
MWC\,623-a & B4\,III & 17200$\pm$3000& 3.7$\pm$0.4& 2.4 & 0.8$\pm$0.2& 7,{\it 7}
&50, {\it 50.0-51.4} & (3)\\
MWC\,623-b & K2\,Ib-II & 4300$\pm$200 & 3.5$\pm$0.4 & 2.4& &
7.5 & 50& (3)\\ 
MWC\,623-b & {\it K4\,I-II} & {\it 4030$\pm$100} & {\it 3.6} & & & {\it 7} & {\it 52} & this work\\
AS\,381-a  & B1& 19000 & 4.9& 4$\pm$1& 2.3$\pm$0.3 & 19$\pm$3, {\it 16}& {\it$\sim$13} & (4)\\
AS\,381-b & K &  & 3.6&4$\pm$1& &7 & & (4)\\
AS\,381-b & {\it K0\,I-II} & {\it 4550$\pm$100}& {\it 3.94$\pm$0.10} & & &\textit{10} & \textit{25}  & this work\\
       \hline
         \end{tabular}

\smallskip
\raggedright{
Parameters given in italic are from this work, others according to the
indicated references: (1)~\citet[distance and $E(B-V)$]{Carmona+2010},
\citet[luminosity]{Miro+1998}, (2)~\citet{Esteban-Fernandez1998},
(3)~\citet{Zickgraf2001}, (4)~\citet{Miro+2002b}, (5)~\citet[spectral
  type]{Hynes+2002}, (6)~\citet{Miro+2002}, (7)~\citet{Miro+2004},
 (8)~\citet[$T_\mathrm{eff}$ and luminosity]{Lobel+2013}

$^a$ Age estimates derived by comparison with single star evolution
models \citep{Ekstroem+2012}, see Fig.\,\ref{fig:hrd} and
discussion of individual stars for further details.}

   \end{table*}

\paragraph*{MWC\,84} Also known as Cl\,Cam, MWC\,84 had a spectacular
outburst in March 1998, detected from $\gamma$-rays to radio emission
\citep[e.g.][]{Clark+1999,Robinson+2002}.  \citet{Hynes+2002}
classified the star as spectral type B0-B2, finding its emission-line
spectrum typical for a B[e]SG. The
authors refer to the star as an atypical high-mass X-ray binary but
speculations about the existence and nature of the compact companion
are ongoing.  \citet{Hjellming+1998} describe the star as X-ray
binary/ X-ray transient radio source, and detect a slow, decelerating
shell in radio emission. However, the nature of this emission is also
still under debate \citep[e.g.][]{Rupen+2003}.  In addition, a faint
circumstellar shell has been detected by \citet{Marston-McCollum2008},
but the authors find no clear evidence to determine whether this is
stellar ejecta or the illuminated local interstellar material.

The analysis of spectral line profiles, the derived extinction, and
interferometric observations suggest that MWC\,84 has an equatorial
disk wind with a dust-free high-temperature zone close to the star,
and that it is viewed almost pole-on
\citep[e.g.][]{Thureau+2009,Hynes+2002,Miro+2002}. The latter authors
revised the distance estimate based on $RV$ measurements and
interstellar \ion{Na}{i} D-lines, to about 2.2\,kpc, placing the star
in the Perseus arm; with that distance, an upper-limit luminosity of
$\log{L/L_\odot} < 4$ is derived. Other studies find both smaller
\citep[$d=1.1-1.9$\,kpc by][$\approx 2\,$kpc]{Barsukova+2006,
  Clark+2000} and larger \citep[e.g. $d \approx
  5\,$kpc][]{Robinson+2002} distances, hampering reliable estimates of
  the luminosity.

\citet{Clark+1999} present $J, H$ and $K$-band spectra obtained one
month after the outburst that are rich in hydrogen, helium and iron
emission lines and clearly show lines of the Pfund series and the
presence of CO emission in the $K$-band. The authors argue that the
molecular emission likely arises due to collisional excitation from
regions shielded from the stellar radiation, which requires high
densities.  
In addition, \citet[][covering 1998 to 1999]{Clark+2000} and
  \citet[][data for 2004 and 2005]{Thureau+2009} present $UBVRIJHK$
  photometry showing that the star is slightly brighter than in pre-outburst
  state, but rather stable over the long time range of seven
  years. Also near-infrared interferometric data \citep[covering 1998
    to 2006]{Thureau+2009} have supported the scenario of a stable
  circumstellar disk in the last decade.

Our spectra (see Fig.\,\ref{fig:mwc084-co}) do not show any
indication of Pfund lines or CO emission or absorption, neither in
2010 (LBT-Luci\,I) nor in 2011 or 2013 (both Gemini-GNIRS).  We
attribute this disappearance of Pfund and CO emission to the
loss/dilution of the high-density circumstellar material and the
return of MWC\,84 to its pre-outburst stage. 

\subsection{Unclassified B[e] stars}
\label{sec:binary}
For both MWC\,623 and AS\,381, it has been suggested that the
stars are in binary systems with a cool companion. Our finding of
prominent CO absorption bands supports this position and we determined
the spectral types for the companions from the CO band head
equivalent widths, see Sect.\,\ref{sec:results} and Table\,\ref{tab:linelist}.

\paragraph*{MWC 623}
Based on the detection of a set of early-type (emission) and late-type
(absorption) optical lines, \citet{Zickgraf-Stahl1989} conclude that
MWC\,623 is a spectroscopic binary (SB2) of type B2+K2. However, they
find no indication for $RV$ variations, attributing this to a binary
period longer than covered by their range of observational data of
2\,years. \citet{Zickgraf2001} reclassifies the spectral types of the
binary components to B4\,III+K2\,II-Ib but no periodic
  $RV$ variations are found even in long-time observations
  \citep{Zickgraf2001, Polster+2012}. This might indicate a pole-on
  orientation of the system, e.g. \citet{Miro2006}.

\begin{figure}
 \includegraphics[width=\columnwidth]{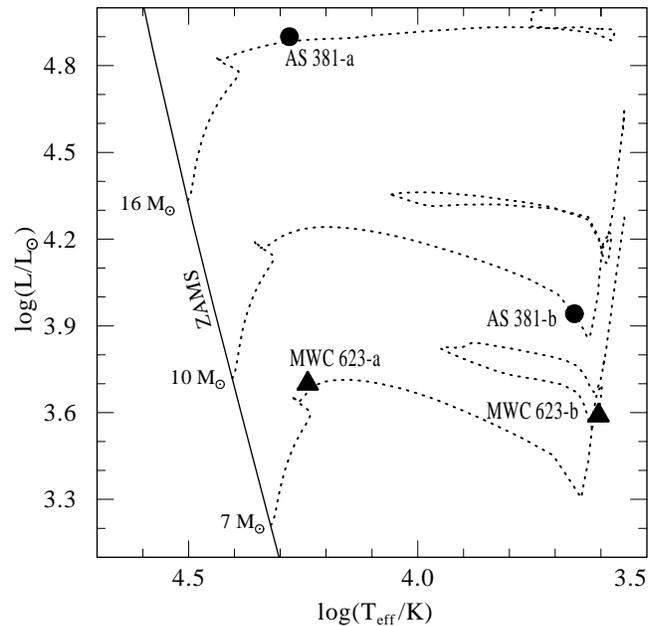}
\caption{HRD for MWC\,623 and AS\,381 and their cool companions;
  single star stellar evolution models for different
  initial masses by \citet{Ekstroem+2012} accounting for the effects
  of stellar rotation.}
  \label{fig:hrd-companions}
\end{figure}

Based on the K star's luminosity class, \citet{Zickgraf2001} derives a
spectroscopic distance of $d =2.4$\,kpc.  Our derived spectral type of
the cool component is a slightly cooler K4\,I-II star, but with the
same range for the luminosity class.  We take MWC\,623's 2MASS
magnitude $K_\mathrm{S} = 5.38$\,mag and assume that the $K$-band
emission is dominated by the cool companion. With the distance and
reddening as listed in Table\,\ref{tab:starparameters}, and bolometric
corrections determined following \citet{Levesque+2005}, we derive the
companion's luminosity as $\log{(L/L_\odot) = 3.61}$.
Assuming that MWC\,623 is a physical long-period binary and not a
chance superposition detected in the spectrum, each component in all
likelihood has evolved like a single star. From comparison with the
stellar-evolution track of 7\,$M_\odot$ initial mass (see
Fig.\,\ref{fig:hrd-companions}), we find an age difference of 1.8\,Myr
of the components. This seems negligible compared to the total age of the
system of about 50\,Myr.

\paragraph*{AS\,381}
The star was first listed in the H\,$\alpha$ surveys
by \citet{Merrill-Burwell1950} and \citet{Henize1976} as emission-line
star; \citet{The+1994} classify it as Be star with IR
excess. \citet{Miro+2002b} detected absorption lines of neutral metals
in their near-IR spectra and $^{12}$CO absorption bands, concluding
that the star is a B1 + K binary system. The authors also derive a
lower limit for the orbital period of $P \approx 30\,$d. They estimate
the distance to be $d = 4 \pm 1$\,kpc deriving luminosities of
$\log{(L/L_\odot)}= 4.9$ and $\log{(L/L_\odot)} = 3.6$ for the B and K
component, respectively. In addition, the authors claim to find signs
of ongoing mass transfer between the binary
components. \citet{Miro2007} lists an effective temperature of
$\log{(T_\mathrm{eff}/\mathrm{K})} = 4.28$ for the B star and an
interstellar reddening of $E(B-V) = 2.2$\,mag.

As described in Sect.\,\ref{sec:results}, we derive a spectral type of
K0 for the companion from the equivalent width measured for the first
$^{12}$CO bandhead. However, the determination of the luminosity class
is not that unambiguous, as a clear detection of $^{13}$CO
is not confirmed.

Assuming that the flux in the infrared-wavelength range is dominated
by the cool companion, we use the mean $K$-band magnitude of those
listed by \citet[][c.f.\ Table\,2]{Miro+2002b}, $K = 6.58\,$mag, to
derive the companion's luminosity. Including the distance and
reddening as listed in Table\,\ref{tab:starparameters}, and the
bolometric corrections for the $K$-band according
to \citet{Levesque+2005}, we obtain $\log{(L/L_\odot) = 3.94 \pm
0.10}$. Within the errors, this result is robust enough to allow up to
25\% of the $K$-band flux to be contributed from the B-type companion.

According to the stellar evolution tracks in
Fig.\,\ref{fig:hrd-companions}, we determine lower-limit initial
masses of about 16\,$M_\odot$ and 10\,$M_\odot$, and ages of about
13\,Myr and 25\,Myr for the B-type and K-type star,
respectively. Fig.\,\ref{fig:hrd-companions} also suggests that the B
component of AS\,381, although more massive, seems to be less evolved
than the K companion.  Given the short lower-limit orbital period of
the system, it seems possible to consider that mass transfer might
have happened at some stage of the evolution of this binary star. In
that case single-star stellar evolution models are of course not
adequate for mass and age determination.

\section{Conclusion}
\label{sec:conclusion}

We conducted a mini-survey of Galactic B[e] stars with the LBT-Luci\,I
and Gemini-GNIRS to characterize their near-infrared $K$-band
spectra. The most dominant emission line feature detected is the
Br\,$\gamma$ line that is present in all our sample stars. In many
cases, iron and magnesium emission can be identified, while helium and
sodium emission lines are less common.  In addition, the detection and
analysis of molecular lines like from carbon monoxide is a very
helpful and powerful tool for the classification of the stars and
their circumstellar material, or for the confirmation of a cool
companion respectively.

Summarizing the results of our B[e] survey:
\begin{enumerate}
\item 
    MWC\,314 shows a spectrum rich in lines of hydrogen,
    including pronounced lines of the Pfund series. Several lines of
    sodium, magnesium, and iron are present as well. Only a weak IR
    excess is detected and CO emission is lacking at all. The $K$-band
    spectrum of MWC\,314 strongly resembles the ones of LBV stars and
    candidates, e.g.\ S\,Dor and LHA\,120-S\,127. Tracers for
    circumstellar material indicate the presence of a non-continuous
    gas disk, i.e.\ rings, around MWC\,314. 
    We determine a spectral type of B2 and an age of about
    6\,Myr; with a lower-limit initial mass of 40\,$M_\odot$ it's the
    most luminous and most massive star in our sample.

\item We detect $^{13}$CO bands in emission in the spectrum of
  MWC\,137 indicating an evolved nature of the star.
  However, the classification of MWC\,137 as Galactic
  supergiant B[e] star has to be confirmed with more data.

\item MWC\,84 shows prominent \ion{He}{i} line emissions. After an
  outburst event in 1998, circumstellar material was clearly and
  stable detected over decades, e.g.\ Pfund lines and CO emission.  We
  observed the star at different epochs with LBT-Luci\,I and
  Gemini-GNIRS and no signs of a recent prominent eruption was found
  during 2010 and 2013. Moreover, the observations reveal the
  disappearance of the circumstellar Pfund lines and the CO emission.

\item MWC\,300 is a B[e] supergiant candidate in a binary system. Its
  spectrum does not show CO emission or absorption at the time of
  observation. 

\item CO absorption bands are found in the spectra of MWC\,623 and
  AS\,381. Both stars have been suspected to be binary systems
  previously. Attributing the observed CO absorption to cool
  companions, we find a spectral classification of B4\,II + K4\,I-II
  (MWC\,623) and B1 + K0\,I-II (AS\,381) and derive the fundamental
  stellar parameters of the companions.

\end{enumerate}

All sample stars are slightly evolved (ages of 6 to 50\,Myr)
with progenitor masses in the intermediate to high-mass range (7 to
40\,$M_\odot$). In the cases of binary stars and candidates,
episodes of mass transfer might have to be considered in the evolution
of the components, thus hampering the exact characterization of the
systems. Future observations, for example radial velocity variations,
pronounced variability, and higher spectral resolution might be needed
to ultimately confirm the binary companions.

Particularly, MWC\,84 and MWC\,300 might be considered as binary
supergiant B[e] candidates in a quiescent transition phase, showing
very low densities in their circumstellar envelopes to sustain
molecular lines. Unfortunately, some supergiant B[e] stars cannot be
distinguished as unambiguously as expected
\citep{Oksala+2013} and further observation are needed to
clarify the evolutionary status of these stars.

\section*{Acknowledgments}

The authors thank Jochen Heidt for useful discussions about planning
the observations and valuable help in preparing the observing scripts;
Steve Allenson and the LBTO team for the support during the observing
runs at the LBT. We thank our referee A. Miroshnichenko for
  valuable comments that helped to improve this manuscript.

AL and AK receive(d) financial support from the Max-Planck-Institut
f\"ur Radioastronomie.  MK acknowledges financial support from
GA\v{C}R under grant number 14-21373S.  The Astronomical Institute
Ond\v{r}ejov is supported by the project RVO:67985815. MK, MLA, and
LSC acknowledge financial support for International Cooperation of the
Czech Republic (M\v{S}MT, 7AMB14AR017) and Argentina (Mincyt-Meys
ARC/13/12 and CONICET 14/003).  MLA and LSC acknowledge
financial support from the Agencia de Promoci\'on Cient\'{\i}fica y
Tecnol\'ogica (Pr\'estamo BID, PICT 2011/0885), from CONICET (PIP
0300), and the Programa de Incentivos G11/109 of the Universidad
Nacional de La Plata, Argentina.

This publication makes use of data products from the Two Micron All
Sky Survey, which is a joint project of the University of
Massachusetts and the Infrared Processing and Analysis
Center/California Institute of Technology, funded by the National
Aeronautics and Space Administration and the National Science
Foundation.

\bsp

\label{lastpage}

\end{document}